\documentclass{article}
\usepackage{ijcai22}

\usepackage{times}

\usepackage{soul}
\usepackage{url}
\usepackage[hidelinks]{hyperref}
\usepackage[utf8]{inputenc}
\usepackage[small]{caption}
\usepackage{graphicx}
\usepackage{amsmath}
\usepackage{booktabs}
\usepackage{notes2bib}

\title{Explanation as Question Answering based on a Task Model of the Agent's Design }

\author{
Ashok Goel$^1$$^2$\and
Harshvardhan Sikka$^1$\footnote{Contact Author}\and
Vrinda Nandan$^1$\and
Jeonghyun Lee$^2$\and\\
Matt Lisle$^2$\And
Spencer Rugaber$^1$\\
\affiliations
$^1$Design \& Intelligence Laboratory
$^2$Center for 21st Century Universities, Georgia Institute of Technology\\
\emails
ashok.goel@cc.gatech.edu,
\{harshsikka, vrinda, jonnalee\}@gatech.edu,
mlisle@gmail.com,
spencer@cc.gatech.edu\\
}

\begin{document}

\maketitle

\begin{abstract}
We describe a stance towards the generation of explanations in AI agents that is both human-centered and design-based.  We collect questions about the working of an AI agent through participatory design by focus groups. We capture an agent’s design through a Task-Method-Knowledge model that explicitly specifies the agent’s tasks and goals, as well as the mechanisms, knowledge and vocabulary it uses for accomplishing the tasks. We illustrate our approach through the generation of explanations in Skillsync, an AI agent that links companies and colleges for worker upskilling and reskilling. In particular, we embed a question-answering agent called AskJill in Skillsync, where AskJill contains a TMK model of Skillsync’s design.  AskJill presently answers human-generated questions about Skillsync’s tasks and vocabulary, and thereby helps explain how it produces its recommendations. 
\end{abstract}

\section{Introduction}

AI research on transparency and explanation faces a familiar conundrum. On one hand, the more complex the design of an AI agent, the larger is the need for making the inner working of the agent transparent to the user. On the other hand, the more complex the design, the more difficult it is to generate explanations of how the agent produces an output. AI faces this conundrum irrespective of the paradigm for designing the AI agent: knowledge-based or data driven, symbolic or connectionist, embodied or software, or some combination of them.  
 
We adopt a stance towards the generation of explanations in AI agents that is both design-based and human-centered. Our stance towards explanation is design-based in that we seek to answer questions about the AI agent based on an explicit model of how the agent produces its results. In particular, we capture the agent’s design through a hierarchical Task-Method-Knowledge (TMK) model that specifies the agent’s tasks and goals, as well as the mechanisms, knowledge and vocabulary it uses for accomplishing the tasks \cite{murdock2008meta,goel2017gaia}. Our stance is human-centered in that the questions are acquired from data collected from humans and the answers are meant for consumption by real users. The questions are acquired through participatory design \cite{muller2012participatory,spinuzzi2005methodology} in focus groups involving key stakeholders.

We present our approach by illustrating the generation of explanations in Skillsync \cite{robson2022intelligent}, an AI agent that links companies and colleges to facilitate worker upskilling and reskilling. In particular, we embed a question-answering agent called AskJill in Skillsync, where AskJill contains a TMK model of Skillsync’s design. AskJill is based on the Jill Watson question-answering technology developed in earlier work \cite{goel2018jill,goel2021explanation}. At present, both Skillsync and AskJill are under development. AskJill presently answers questions only about Skillsync’s tasks and vocabulary. Our research hypothesis is that a TMK model of an agent’s design both provides a scheme for classifying questions asked by users and captures the knowledge needed to answer questions that human stakeholders typically ask about the working of the agent. We present preliminary results on the evaluation of AskJill in Skillsync.

\section{Background}

\subsection{Skillsync}
\begin{figure}[ht!]
    \centering
    \includegraphics[width=\linewidth]{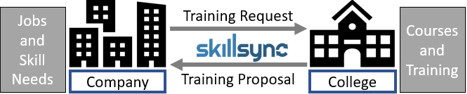}
    \caption{An overview of Skillsync, adapted from \protect\cite{robson2022intelligent}}
    \label{fig:Skillsync}
\end{figure}

The Skillsync application \cite{robson2022making,robson2022intelligent} helps companies address the need to reskill or upskill their employees in partnership with colleges. It also helps colleges match their continuing education and professional development programs to the needs of industry. Figure \ref{fig:Skillsync} illustrates the workflow in Skillsync. The application enables companies to document the needed skills in the form of training requests and send these training requests to relevant education providers. It also allows colleges to formulate training proposals in response to the requests based on their educational programs. Skillsync uses various AI techniques, including machine learning, language models, and matching algorithms, to extract knowledge, skills, and abilities (KSA) \cite{cdc,usva}. % need to add citation /\cite{cdc,usva}/  with dates
from job data and to match them with corresponding courses. The job data originates from sources like the U.S. Department of Labor, industry associations, company job descriptions, and job postings provided by the National Labor Exchange (https://usnlx.com/).  The course data is sourced from course catalogs of continuing education and professional development programs at universities and colleges, including technical and community colleges. The extracted KSAs are organized and prioritized in a skills framework. Skillsync helps make the process of matching jobs with educational programs both more efficient and effective. However, the potential adoption and use of Skillsync in companies and colleges requires that its results are trustworthy, that its processing is transparent, and that it can explain its design and processing. 

\subsection{TMK Models}
Task-Method-Knowledge (TMK) modeling constitutes a formal approach to building machine readable knowledge representations of AI agents with the ultimate goal of empowering the agents to reason about other agents or about themselves. By interacting with a TMK model, an AI agent is able to provide causal explanations and manipulate various aspects of the agent being modeled \cite{goel1996explanatory}. TMK models also fulfill a separate but related goal: providing a machine readable formalism that is also interpretable by humans interacting with the AI agent in question. TMK models encode information in three different ways: Tasks represent the \textit{why} of a system, specifying the goals of the AI agent. Methods represent the \textit{how} of a system, describing the internal processing of the agent. Knowledge captures the \textit{what} of the AI agent, expressing the information that the agent is operating on. TMK models are compositional, causal, and hierarchical. In particular, a method for a task decomposes the task  into subtasks that have methods of their own. The tasks at the leaf level are directly accomplished by a chunk of knowledge, an action in the agent, or an interaction with a user. TMK models have been applied to multiple domains ranging from navigational and assembly planning to design of game-playing agents. AskJill in Skillsync uses a TMK model of Skillsync to generate explanations of how it works.

\subsection{Jill Watson}

AskJill has evolved from the virtual teaching assistant AI agent, Jill Watson. The goal of the Jill Watson QA agent is to amplify teacher presence in online learning environments. Jill Watson was initially developed in the context of the Online Master of Science in Computer Science (OMSCS) program at Georgia Tech launched in 2014. Students in the program often interact with instructors through discussion forums. Classes in the program demonstrated the need for amplifying teacher presence quickly following launch, as students posted hundreds of questions in the associated discussion forums, resulting in difficulties for the instructional team to successfully answer them all. Jill Watson uses homegrown technology on top of IBM’s Watson platform \cite{ferrucci2010building}. Conceptually, the Jill Watson Q \& A agent uses a hybrid classification approach to answer questions pertaining to the syllabus of a course. First, Jill Watson uses statistical machine learning methods to categorize the underlying intent behind an incoming question from a student. Subsequently, a knowledge based classifier parses the question and structures an appropriate response from an underlying knowledgebase of relevant course related information. While the original Jill Watson answered students’ logistical questions about a class, more recently, we developed a variation of Jill Watson called AskJill that answers users’ content questions based on a User Guide \cite{goel2022agent}. Jill Watson is now being adapted to provide explanations for systems and agents like Skillsync.

\section{ Participatory Design, Question Classification, and Requirement Analysis}

\subsection{Focus Groups and Question Generation}
Guided by Spinuzzi's~\shortcite{spinuzzi2005methodology} methodology, our research focused on the initial exploration and discovery stages of participatory design involving walkthroughs, user observations, and feedback solicitation. Specifically, most of the data came from user comments and answers to guided prompts that were designed to elicit user feedback regarding their tasks and needs. A total of five focus groups were conducted between September and December 2020 to gather feedback from potential users of Skillsync through the process of collective discovery 
\cite{muller2012participatory}. By using data generated from these focus groups, we aimed to draw inferences about the possible usage of AskJill and corresponding user requirements. Specifically, we focused on identifying notable questions or issues that participants addressed while they were interacting with the Skillsync prototype.
 
Across the focus group sessions, there were fifteen participants representing ten employers (i.e., company users) and five educators (i.e., college users). Participants were invited to a breakout activity in which the focus group facilitator showed a demonstration of how a company requests a training project and based on this information how a college builds and shares a proposal with the company. When interacting with Skillsync, company users were asked to provide detailed descriptions of target trainees (e.g., current skills or competencies) and training that they would request from a college provider; college users were asked to review a training request submitted from a company user and then build a training proposal that includes a list of courses that the college can offer to best meet the training needs.
 
Participants watched the facilitator or other participants navigating the prototype through a shared screen. The facilitator paused at several landing pages to elicit participants’ feedback about what questions they would ask of AskJill regarding how to use the Skillsync interface. Participants were encouraged to share any questions that arose as they explored the prototype. The primary goal of this question-generation activity was to uncover salient types of questions that users would ask of an AI agent. As a result, we were able to extract 52 questions that were explicitly raised by the participants, which became the basis of our question classification task.

\begin{figure}[ht!]
    \centering
    \includegraphics[width=1.025\linewidth]{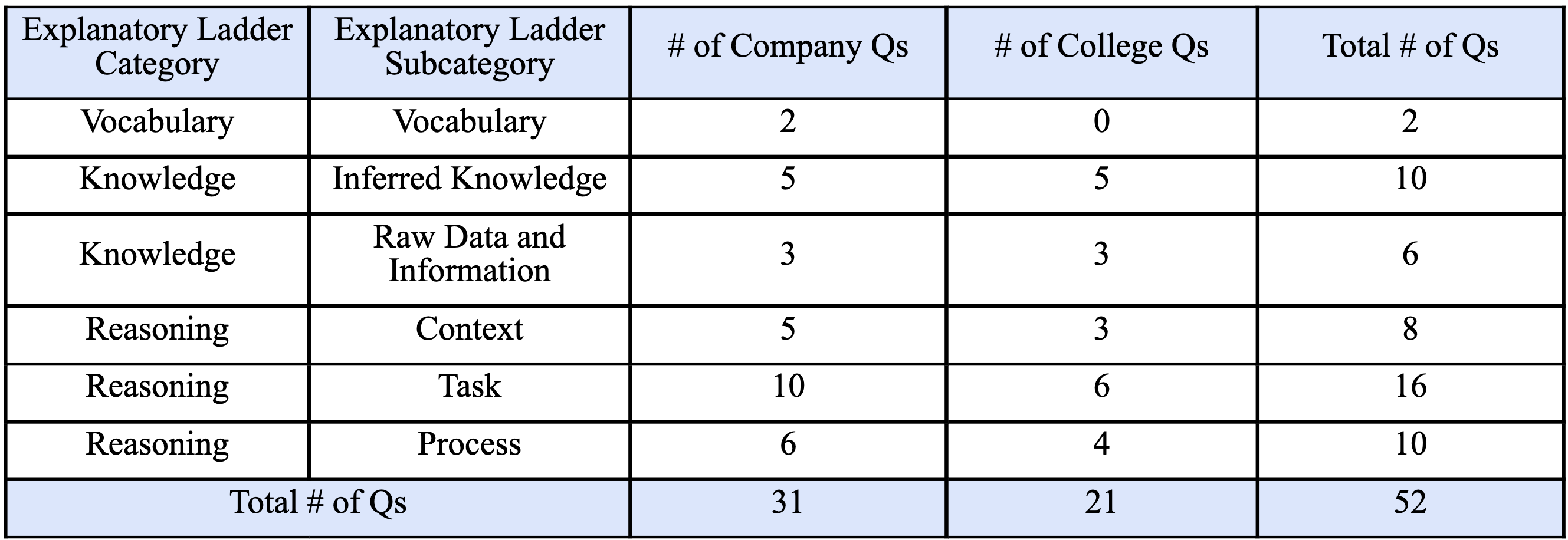}
    \caption{Classification of Questions from Companies and Colleges.}
    \label{fig:participatory_questions}
\end{figure}

Explanations in Skillsync help build a shared vocabulary and mental model between AI and the company and college professionals. In order to inform development and training of the knowledge base for AskJill, user questions that we extracted from the focus group data were categorized based on an Explanatory Ladder derived from the TMK framework. Categories and subcategories from the explanatory ladder are included in Figure \ref{fig:participatory_questions}. According to this Explanatory Ladder, there are three main categories of explanations, where the depth of explanations increases as we go up the ladder. At the lowest level of the ladder, the agent can answer questions about its vocabulary; at the next level, questions about knowledge in two subcategories  (raw data and information, and inferred knowledge); and the highest level, questions about the reasoning with three subcategories (context of the current interaction, task, the process to accomplish those tasks). 

% explanatory ladder figure is not included for space
% \begin{figure}[ht!]
%     \centering
%     \includegraphics[width=1.025\linewidth]{image4.png}
%     \caption{Explanatory Ladder for Skillsync}
%     \label{fig:explanatory_ladder}
% \end{figure}

Of the 52 user questions that we extracted, 31 questions were relevant to company users and the remaining 21 to college users. The results of the question classification in the Explanatory Ladder are shown in Figure \ref{fig:participatory_questions}. Our classification results suggest that both types of users actively generated task-related questions that sought clarification or guidance for task completion (e.g., \textit{How do I add a competency?} \textit{How do I select and add occupational tasks?}).

Overall, our participatory design study led to eliciting authentic and quality feedback. Most focus group participants will be actual users of the interface in the future. Therefore, they tended to be highly interested and motivated to understand the values of the Skillsync interface or how it would meet their specific needs. By using the question-generating activity, we had an opportunity to explicitly bring their awareness and attention to the AskJill tool prior to the pilot experiment. This allowed us to gain knowledge on what comes to a user’s mind and issues that need to be addressed.
 
\section{ A TMK Model of Skillsync}

\subsection{Specification of Skillsync’s Vocabulary}

\begin{figure}[ht!]
    \centering
    \includegraphics[width=1.1\linewidth]{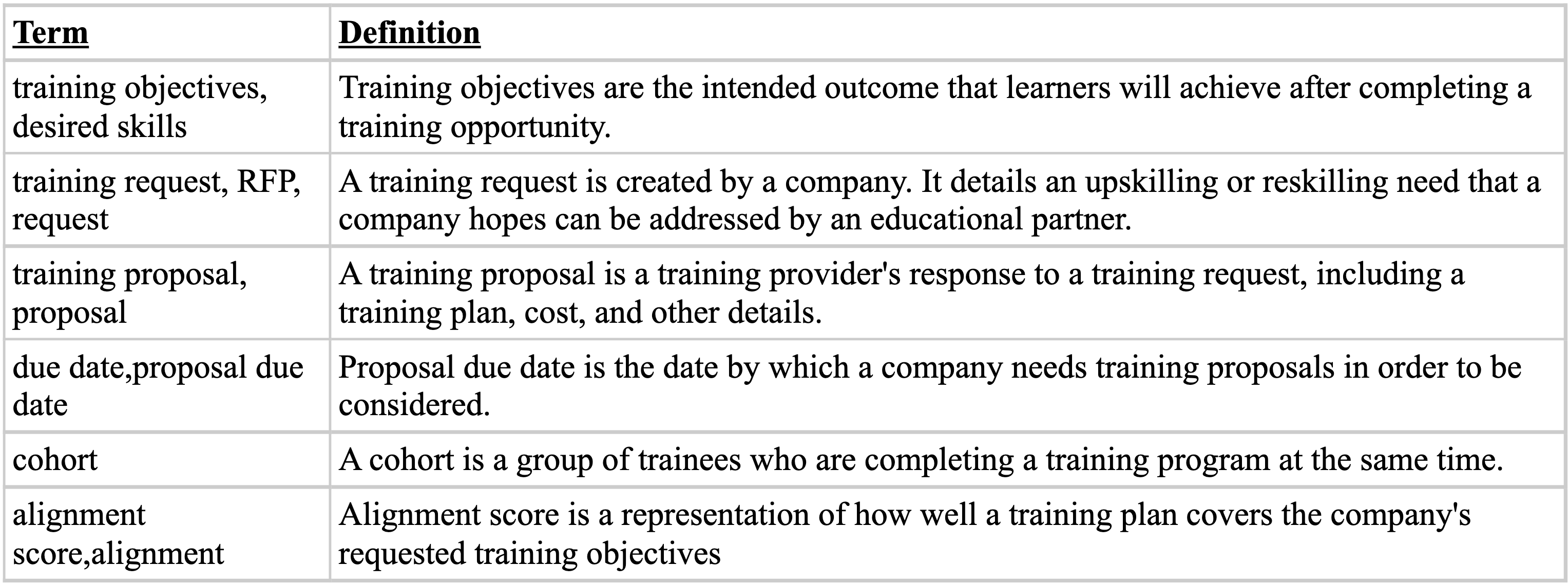}
    \caption{A subset of Skillsync's Vocabulary}
    \label{fig:vocabulary_terms}
\end{figure}

The TMK model of an agent specifies the vocabulary used by the agent. Figure \ref{fig:vocabulary_terms} includes a representative list of AskJill’s knowledge of Skillsync’s vocabulary (terms and their definitions). In total, we have 41 terms and definitions in AskJill’s knowledge base. 

\subsection{A Task Model of Skillsync}

AskJill’s task model of Skillsync comprises task goals, subtasks, and the inputs and the outputs of the task. This information is stored in AskJill’s knowledge base and captures a hierarchy of 10 tasks that the user can accomplish on Skillsync, from high-level tasks to primitive actions. For example, to accomplish the task of creating a training plan, the company user needs to create and submit a training request and the college user needs to respond with a well-aligned training proposal. The two users can then agree to move forward with the training, thereby accomplishing their goal of creating a joint training plan. The task model hierarchically breaks down the tasks related to the two tasks until it reaches the primitive tasks. The lowest levels in the task hierarchy involve primitive user actions such as button click, text entry, and file upload. Table \ref{tab:tmk} captures the “training request” and “training proposal” tasks as examples, showing the associated  task goals, subtasks, inputs and outputs. 

%need to creatively adapt tmk figure from section 4, too large and unreadable currently. 

{\renewcommand{\arraystretch}{1.3}%
\begin{table*}[ht!]
    \centering
    \scriptsize
    \begin{tabular}{|p{0.1\linewidth}|p{0.4\linewidth}|p{0.4\linewidth}|}
        \hline 
        \textbf{Task Keywords} & Training Request, RFP, Request & Training Proposal, Proposal \\ 
        \hline 
        \textbf{Goals} & A training request is created by a company. It details an upskilling or reskilling need that a company hopes can be addressed by an educational partner. & College creates a training proposal that meets the needs detailed in the company partner’s transmitted training request. \\ 
        \hline 
        \textbf{Inputs} & College creates a training proposal that meets the needs detailed in the company partner’s transmitted training request. & Receive the transmitted Training Request, Review Training Opportunities Catalog, Select Training Opportunities, Enter Proposal Details, Create Training Proposal Summary  \\ 
        \hline 
        \textbf{Outputs} & Completed Training Request & Alignment Score, Completed Training Proposal \\ 
        \hline
    \end{tabular}
    \caption{Examples tasks from TMK.}
    \label{tab:tmk}
\end{table*}}

\section{ AskJill, A Question-Answering Agent for Skillsync}

\subsection{An Overview of AskJill}
AskJill is a question answering agent with the goal of providing explanations for tasks and vocabulary present in the Skillsync Platform. When a user logs into Skillsync, AskJill can be accessed via a text window. The user can type their question and can expect precise answers from AskJill within seconds. Figure \ref{fig:askjill_screenshots} demonstrates AskJill answering user questions on the Skillsync Platform. In this section, we outline the architecture and underlying concepts that make up the AskJill system, including how incoming questions are classified and parsed, how answers are generated, and how AskJill agents are trained to do all of the above.

\begin{figure}[ht!]
    \centering
    \includegraphics[width=1\linewidth]{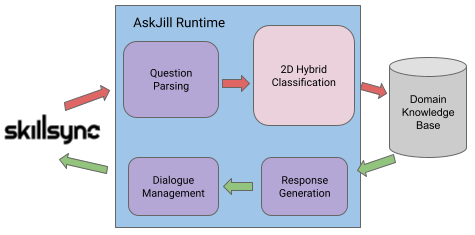}
    \caption{AskJill Dataflow, adapted from \protect\cite{goel2022agent}}
    \label{fig:data_flow}
\end{figure}

\subsection{Data Flow of an Answer to a Question}
When a user asks a question on the Skillsync platform, the question is sent via a REST API to the AskJill system. Once received, the question is parsed and then sent to 2D hybrid classification system. In 2D classification, AskJill makes use of two approaches in tandem: a natural language intent classifier that is trained using machine learning and a set of semantic rules that process classified intents into structured queries. Following classification, these queries are sent to the Skillsync knowledge base, where a response is generated based on querying the domain knowledge base. With an appropriately high confidence exceeding a tuned threshold, the response is passed through the dialogue management system, which converts the answer to a natural language response styled after humanlike conversation. This response is sent back to the Skillsync system and displayed on the text interface. After answering, AskJill prompts the user to provide feedback, asking “Was this answer helpful”, and stores the user feedback in its database. That feedback is subsequently used for retraining the agent. If AskJill is unable to answer a question, it can gently redirect the conversation to its domain of competence by suggesting alternate topics associated with the questions it is trained on. Figure \ref{fig:data_flow} includes an overview of the data flow and hybrid classification step when the AskJill runtime answers a question.

\subsection{Answering Example Questions}
\begin{figure*}[ht!]
    \centering
    \includegraphics[width=1\linewidth]{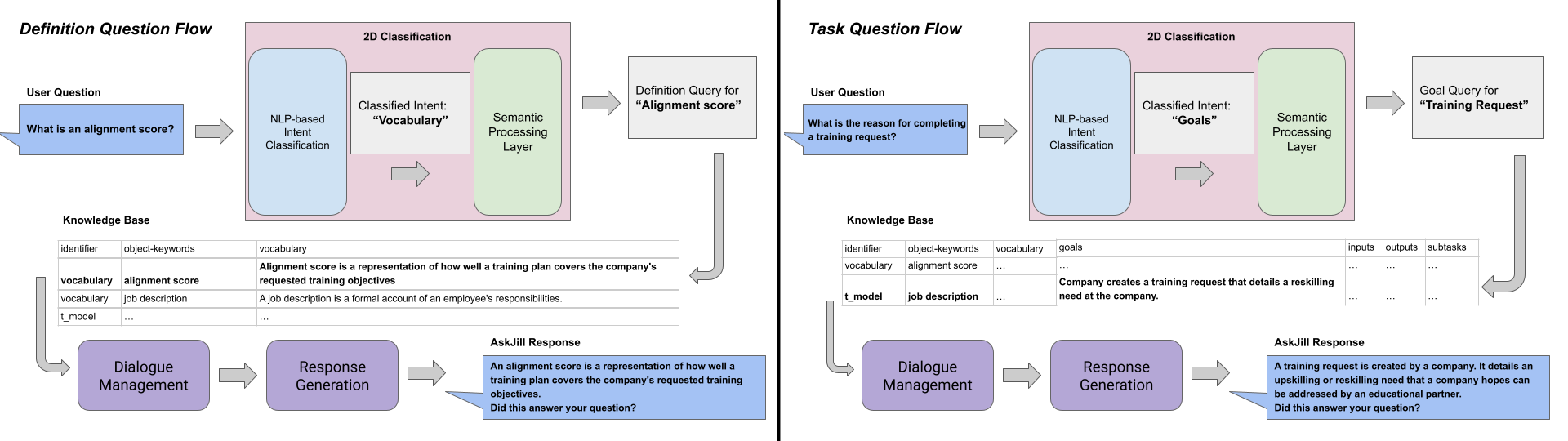}
    \caption{Examples of AskJill QA process for Vocabulary and Task related questions.}
    \label{fig:question_flow}
\end{figure*}

Let us now consider a couple of examples. First, say a user asks, “\textit{What is an alignment score?}” via the AskJill text interface on the Skillsync website. AskJill parses the question and sends it to the 2D classification system, where the first layer, the natural language intent classifier, determines that the question relates to vocabulary. Then, the second layer, the rule based classifier, creates a structured query referring to “alignment score”. The query is sent to the Domain Knowledge base, implemented in MySQL, that holds the definitions for various terms in the Skillsync glossary. AskJill retrieves the definition of “alignment score” from the knowledge base and formats its response, “Alignment score is a representation of how well a training plan covers the company's requested training objectives”. This response is formatted into natural language and sent to the user via the same text window. Similarly, when a user asks “\textit{What is the reason for completing a training request?}”AskJill extracts the intent as “goals” and retrieves the goal associated with a training  request. In that case, AskJill responds, “A training request is created by a company. It details an upskilling or reskilling need that a company hopes can be addressed by an educational partner”. In both cases, AskJill requests user feedback, asking “Was this answer helpful?”, and stores the user feedback in its database. Illustrative examples are provided in Figure \ref{fig:question_flow}.

\subsection{Training AskJill with Machine Teaching}
AskJill agents can be rapidly created to support a variety of domains \cite{goel2018jill,goel2021explanation,goel2022agent}, with limited labor and time costs through the use of an interactive machine teaching environment known as Agent Smith. The first step in enabling the Agent Smith system to rapidly train an AskJill Agent is to map the domain to the unstructured and structured databases. As mentioned earlier, the current focus of AskJill in the Skillsync domain is answering questions about Skillsync’s tasks and the vocabulary a user may encounter on the platform.

% need to review the figure requirements here
\subsubsection{Mapping Vocabulary and Tasks to the Knowledgebase}
To capture relevant vocabulary from Skillsync in the AskJill knowledge base, key terms were identified based on results from participatory design studies of potential users outlined in Section 3, as well as the conceptual relevance of the term to the domain. These terms and their associated definitions are mapped to the domain knowledge base, which has two distinct tables: the structured and unstructured table. The former has a set of specific pieces of information associated with different kinds of terms within the broader vocabulary and TMK categories, while the latter contains  text responses associated with the category. These tables are used during the training process, and again by the AskJill agent during runtime when answering user questions. Tasks were mapped in a similar way, with different actions and interactions taken by users mapped to a Task model, also contained in the same structured and unstructured tables contained in the domain knowledge base. % We include the schemas in Figure X. 
% No schema/tmk database figure. We already have a figure for the TMK before that still needs to be included. 

\subsubsection{Capturing the Form of Potential Questions}
The next important step is to collect a set of questions that will likely be asked about Skillsync as outlined in Section 3.1. These questions are then mapped to a general list of template questions and their associated intents. Template questions seek to capture general forms of questions that occur within categories in the domain, representing a large set of potential questions. These templates are designed by extracting common patterns in the participatory design work outlined in Section 3.1, and classifying them as outlined in Section 3.2. In Table \ref{tab:templates}, we demonstrate example template questions for different categories in the explanatory ladder. These templates are stored in their own database.

{\renewcommand{\arraystretch}{1.3}%
\begin{table}[ht!]
    \centering
    \scriptsize
    \begin{tabular}{|p{0.3\linewidth}|p{0.6\linewidth}|}
        \hline 
        
        \textbf{Type} & \textbf{Example Template} \\ 
        
        \hline 
        Vocabulary & What is \{object\}? \\ 
        \hline 
        Inputs (tasks) & What inputs do I need to complete this \{object\}? \\ 
         \hline 
         Goals (tasks) & What is the goal of \{object\}?  \\ 
         \hline 
         Outputs (tasks) & What is the expected outcome of \{object\}? \\ 
         \hline 
         Subtasks (tasks) & What are the steps to accomplish \{object\}? \\ 
         \hline 
    \end{tabular}
    \caption{Examples of training templates used to generate training data.}
    \label{tab:templates}
\end{table}}

\subsubsection{Dataset Generation}
After the creation of the databases underlying the knowledge base, as well as the template questions database, Agent Smith can be used to create a large dataset of example question-answer pairs to train an AskJill agent. This is accomplished through combinatorially connecting template questions with various structured and unstructured keywords that represent vocabulary and TMK related concepts. 
% In Figure X, we demonstrate a few examples of questions generated from some of the templates demonstrated in the previous table, along with their associated intents.
% No example questions figure/table

% currently included agent smith figure, can be removed due to space
% \begin{figure}[ht!]
%     \centering
%     \includegraphics[width=\linewidth]{agentsmith.png}
%     \caption{Overview of Agent Smith Generator}
%     \label{fig:agentsmith}
% \end{figure}

% should the below section be included?
% X questions and intent pairs were created following the running of Agent Smith on the Skillsync knowledge bases and template questions created as of October 2021. These were subsequently used to train the machine learning classifier underlying the 2D classification portion of the AskJill agent.

\section{Evaluation of AskJill in Skillsync}
\begin{figure}[ht!]
    \centering
    \includegraphics[width=1.05\linewidth]{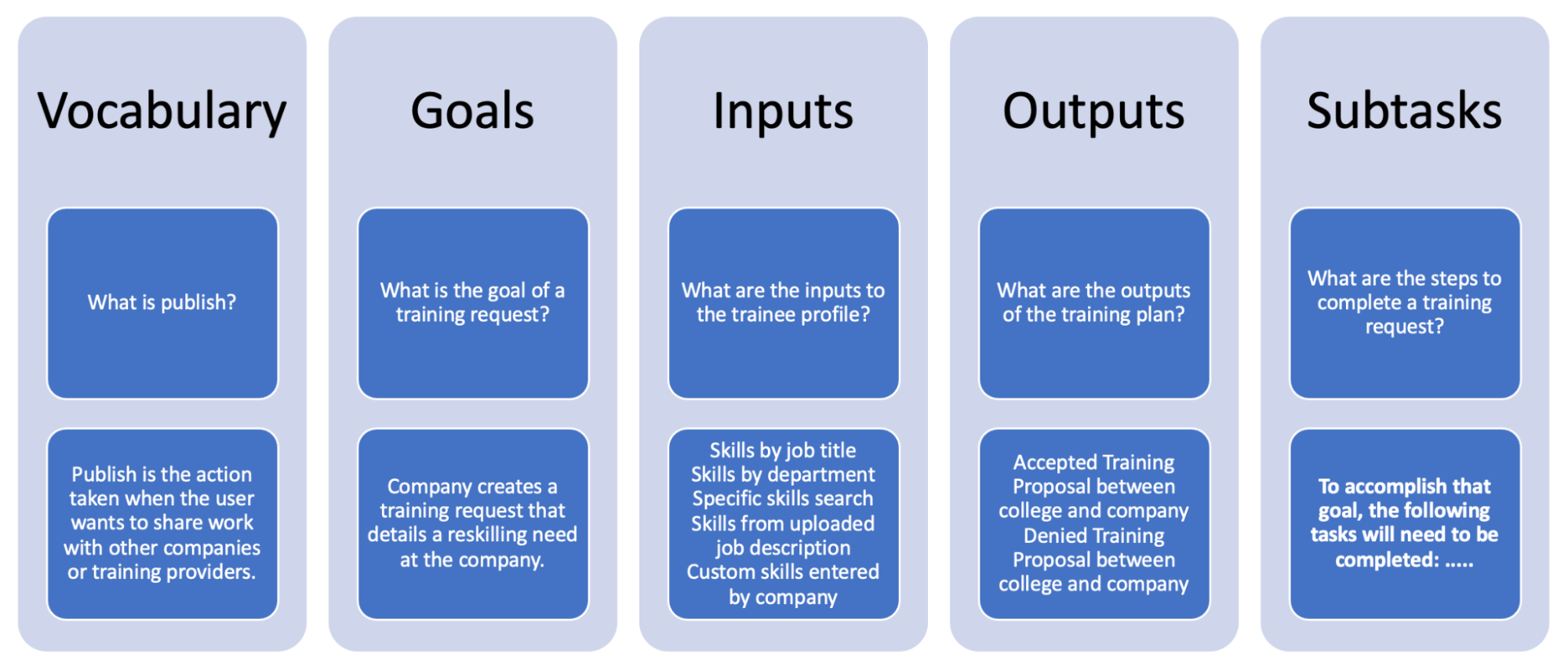}
    \caption{Categories of questions AskJill can answer.}
    \label{fig:categories}
\end{figure}

As part of AskJill’s development process, we compiled a glossary of terms (or vocabulary) and developed a task model for Skillsync. We added this information to AskJill’s knowledge domain and subsequently trained AskJill to answer a large set of human-generated questions pertaining to vocabulary  (relates to “knowledge” in TMK) and users’ tasks in Skillsync. We proceeded to gather both in-vitro and in-situ data from AskJill users between June 2021 and September 2021. This period included Skillsync’s pilot trial, conducted with participants from Skillsync’s partner colleges and companies. During this time, users asked questions about Skillsync’s glossary of terms (vocabulary) and the tasks they can complete on Skillsync (task goals, inputs, outputs, subtasks). Figure \ref{fig:categories} captures the general categories of questions that AskJill can answer. AskJill was able to answer questions that fall into all five categories (or user intents): vocabulary, task goals, inputs, outputs, subtasks. We validated AskJill’s question-answering abilities in-vitro, using its own training dataset of both real user questions and anticipated questions based on known templates. AskJill correctly answered all 1511 questions in the training dataset.

\begin{figure}[ht!]
    \centering
    \includegraphics[width=1.1\linewidth]{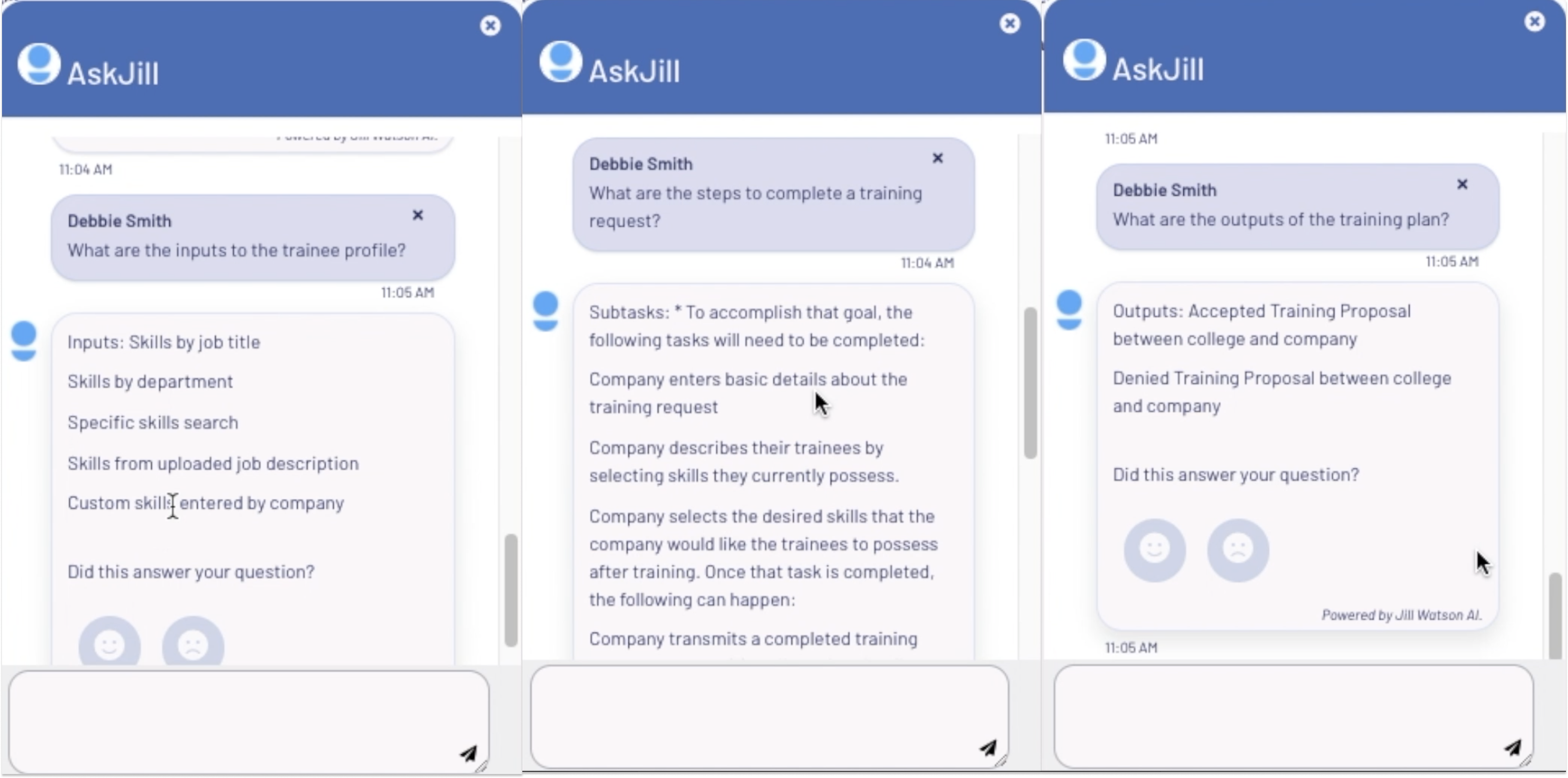}
    \caption{Examples of AskJill's response}
    \label{fig:askjill_screenshots}
\end{figure}

\begin{figure}[ht!]
    \centering
    \includegraphics[width=1.025\linewidth]{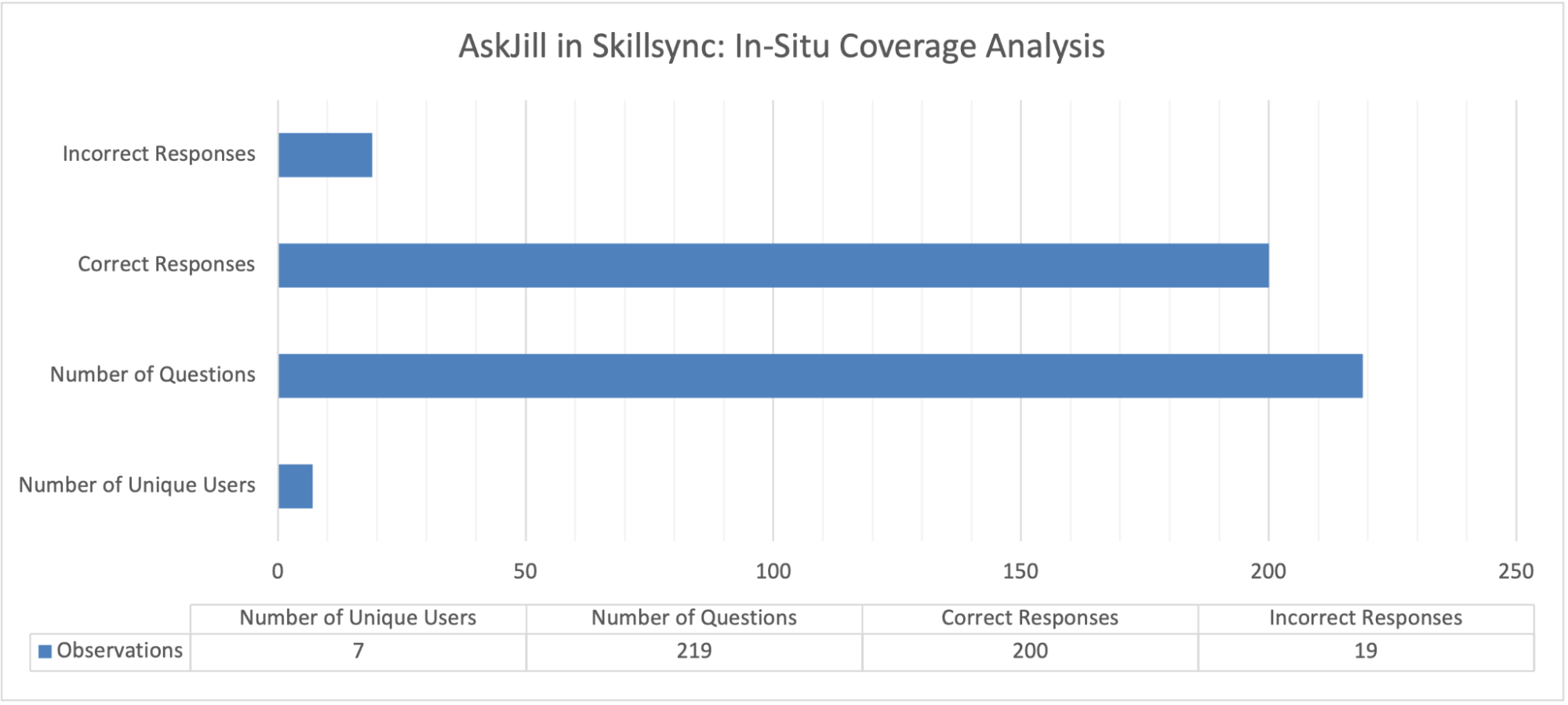}
    \caption{In-Situ Coverage Analysis of AskJill.}
    \label{fig:coverage}
\end{figure}

Figure \ref{fig:askjill_screenshots} shows examples of human generated questions and AskJill’s agent-generated responses to the questions directly from the AskJill text interface on Skillsync’s UI. Notice that AskJill is able to provide answers both vocabulary terms and tasks (goals, inputs, outputs and subtasks) on Skillsync. In addition, we gathered a small dataset of in-situ observations. These observations were collected from college and company users interacting with AskJill, embedded in Skillsync, on the platform’s website. Figure \ref{fig:coverage} depicts a comparison of data collected from seven unique users who interacted with AskJill via a text window embedded in the Skillsync UI. Altogether, they asked 219 questions (of which 106 were unique). We validated that AskJill correctly answered 200 questions (91\%). Out of the 19 missed questions, 1 had a major language error, 10 were outside of AskJill’s competence (e.g. What is the weather today?) and 8 referred to old UI terms that had been removed from the Skillsync UI. We also assessed user satisfaction directly using a feedback question (Did this answer your question?) that is integrated into the agent's type-in box in the UI. We confirmed that the users indeed provided positive feedback to the correct responses (although occasionally users did not share any feedback).

\section{Discussion}
AI research on explanation has a long history that dates at least as far back as the rise of expert systems in the 1960s, e.g., DENDRAL \cite{lindsay1993dendral}. Mueller et al.~\shortcite{mueller2019explanation} provide a recent and comprehensive review of this research. One of the key ideas to emerge out of this early research was the importance of the explicit representation of knowledge of the design of an AI system \cite{chandrasekaran1991explanations,chandrasekaran1989explaining}, which enables generation of explanations of the tasks it accomplishes, the domain knowledge it uses, as well as the methods that use the knowledge to achieve the tasks. This raised the question of how this design knowledge can be identified, acquired, represented, stored, accessed, and used for generating explanations. One possible answer was to endow the AI agent with meta-knowledge of its own design (e.g., \cite{goel1996explanatory}).  

Although much of the research on expert systems collapsed in the mid-nineties, explanation continued to attract attention in some schools of AI such as case-based reasoning \cite{leake2005introduction,schank2014inside} and intelligent tutoring systems \cite{aleven2002effective,woolf2010building}, often in the form of question-answering \cite{graesser1996question}. Over the last few years, explanation has again become important in mainstream AI research (e.g., \cite{gunning2019darpa}). This is in part because of advances in machine learning, such as deep learning, that have refocused attention on the need for interpretability and explainability of internal representations and processing in AI agents. 

Question and Answering covers a large variety of different approaches, and there have been several notable works in Q \& A using knowledge representations. Predefined rule based approaches like Bast et al.~\shortcite{bast2015more} use templates to extract logic. More recently,  Information Retrieval (IR) and Neural Semantic Parser  (NSP) based approaches have been proposed \cite{fu2020survey}. The former focuses on extracting entities from natural language questions and makes connections to the knowledge base being covered. Representation learning has been used to great effect in IR based approaches. These methods map question answer pairings into vector space and formulate a matching problem within the distribution of questions and answers. External knowledge has been introduced in the form of structured knowledge bases, web corpuses and the like. Xu et al.~\shortcite{xu2016question} used Wikipedia as a knowledge base for a KBQA method. Watson \cite{ferrucci2010building}, and thereby AskJill, uses a similar IR representation learning based approach to encode questions into the numerical space of a classifier. 

% Machine teaching and data augmentation are also areas that AskJill touches on. Data augmentation is a large area, but template based approaches have been used to great effectiveness. Linguistic knowledge based approaches have been suggested,  incorporating prior knowledge to form the augmentation, and templates have been used successfully in Q\&A applications prior to this as well \cite{singh2019xlda,asai2020logic}. In Machine teaching (MT), Simard et al.~\shortcite{simard2017machine} focus on improving the efficacy of teachers, emphasizing on the interaction of a teacher with the data, while Zhu et al.~\shortcite{zhu2015machine,zhu2018overview} present a formal perspective on MT, characterizing a problem space that describes the topic. 

In the future, we plan to extend the TMK model so that AskJill can answer user questions related to the knowledge and methods used in Skillsync. Given the significant changes to the Skillsync platform and user interface based on the two user trials over the development period, we also plan to host additional user focus group sessions to extract additional questions about Skillsync. This participatory design approach enables us to maximize AskJill’s question answering ability and enables AskJill to provide explanations for many more real time user generated questions.
Another limitation of the current version of AskJill is that it does not afford explanations of specific instances of reasoning and action by the AI agent. Thus, this approach likely has to be complemented with an episodic approach that relies on specific cases of decision making \cite{langley2017explainable}. In our own earlier work along these lines, we used meta-cases to capture derivational traces in an earlier interactive learning environment and used the meta-cases to explain the agent’s decision making \cite{goel1996explanatory}. A future version of AskJill may keep a derivational trace of Skillsync’s decision making and augment its explanatory capability based on a replay of the derivational trace.

\section{Summary and Conclusions}
A responsible, trustworthy and transparent AI agent must be able to explain how it works and produces its results. Skillsync is an AI agent that links companies and colleges for worker upskilling and reskilling. While Skillsync is useful, its adoption in practice likely will depend in part on its explainability. Thus, we have embedded AskJill, a question-answering agent, into Skillsync. AskJill answers questions about Skillsync’s tasks and vocabulary. From the perspective of human-centered AI, we collected and classified questions about the working of Skillsync through participatory design by focus groups. From the viewpoint of design-based explanation, we captured the design of Skillsync through a task model that explicitly specifies its tasks and goals, as well as the vocabulary it uses for accomplishing the tasks. The main takeaway from this work is the usefulness of an explicit model (TMK) of an agent’s design for (1) classifying users’ questions and  (2) question-answering mechanisms for explaining the agent’s tasks, goals and vocabulary. In future work, we will complete the TMK model of Skillsync so that AskJill can also answer questions about the methods and knowledge Skillsync uses to complete its tasks.

\section*{Acknowledgements}

Our work on AskJill in Skillsync is sponsored by the US National Science Foundation through a Convergence Accelerator grant (\#2033578). We thank our collaborators at Eduworks who developed Skillsync for their help with AskJill including Robby Robson, Elain Kelsey, Kristin Wood, and Alan LaFleur. AskJill in Skillsync uses IBM’s Watson platform for intent classification. We thank IBM for its support for our work. However, the authors alone are responsible for the contents of this paper.

\bibliographystyle{named}
\bibliography{paper}

\begin{thebibliography}{}

\bibitem[\protect\citeauthoryear{Aleven and
  Koedinger}{2002}]{aleven2002effective}
Vincent~AWMM Aleven and Kenneth~R Koedinger.
\newblock An effective metacognitive strategy: Learning by doing and explaining
  with a computer-based cognitive tutor.
\newblock {\em Cognitive science}, 26(2):147--179, 2002.

\bibitem[\protect\citeauthoryear{Bast and Haussmann}{2015}]{bast2015more}
Hannah Bast and Elmar Haussmann.
\newblock More accurate question answering on freebase.
\newblock In {\em Proceedings of the 24th ACM International on Conference on
  Information and Knowledge Management}, pages 1431--1440, 2015.

\bibitem[\protect\citeauthoryear{Chandrasekaran and
  Swartout}{1991}]{chandrasekaran1991explanations}
B~Chandrasekaran and William Swartout.
\newblock Explanations in knowledge systems: the role of explicit
  representation of design knowledge.
\newblock {\em IEEE expert}, 6(3):47--49, 1991.

\bibitem[\protect\citeauthoryear{Chandrasekaran \bgroup \em et al.\egroup
  }{1989}]{chandrasekaran1989explaining}
B~Chandrasekaran, Michael~C Tanner, and John~R Josephson.
\newblock Explaining control strategies in problem solving.
\newblock {\em IEEE Intelligent Systems}, 4(01):9--15, 1989.

\bibitem[\protect\citeauthoryear{Ferrucci \bgroup \em et al.\egroup
  }{2010}]{ferrucci2010building}
David Ferrucci, Eric Brown, Jennifer Chu-Carroll, James Fan, David Gondek,
  Aditya~A Kalyanpur, Adam Lally, J~William Murdock, Eric Nyberg, John Prager,
  et~al.
\newblock Building watson: An overview of the deepqa project.
\newblock {\em AI magazine}, 31(3):59--79, 2010.

\bibitem[\protect\citeauthoryear{Fu \bgroup \em et al.\egroup
  }{2020}]{fu2020survey}
Bin Fu, Yunqi Qiu, Chengguang Tang, Yang Li, Haiyang Yu, and Jian Sun.
\newblock A survey on complex question answering over knowledge base: Recent
  advances and challenges.
\newblock {\em arXiv preprint arXiv:2007.13069}, 2020.

\bibitem[\protect\citeauthoryear{Goel and Polepeddi}{2018}]{goel2018jill}
Ashok~K Goel and Lalith Polepeddi.
\newblock Jill watson: A virtual teaching assistant for online education.
\newblock In {\em Learning engineering for online education}, pages 120--143.
  Routledge, 2018.

\bibitem[\protect\citeauthoryear{Goel and Rugaber}{2017}]{goel2017gaia}
Ashok~K Goel and Spencer Rugaber.
\newblock Gaia: A cad-like environment for designing game-playing agents.
\newblock {\em IEEE Intelligent Systems}, 32(3):60--67, 2017.

\bibitem[\protect\citeauthoryear{Goel \bgroup \em et al.\egroup
  }{1996}]{goel1996explanatory}
Ashok Goel, Andr{\'e}s G{\'o}mez~de Silver~Garza, Nathalie Gru{\'e}, J~William
  Murdock, Margaret Recker, and T~Govindaraj.
\newblock Explanatory interface in interactive design environments.
\newblock In {\em Artificial intelligence in design’96}, pages 387--405.
  Springer, 1996.

\bibitem[\protect\citeauthoryear{Goel \bgroup \em et al.\egroup
  }{2021}]{goel2021explanation}
Ashok Goel, Vrinda Nandan, Eric Gregori, Sungeun An, and Spencer Rugaber.
\newblock Explanation as question answering based on design knowledge.
\newblock {\em arXiv preprint arXiv:2112.09616}, 2021.

\bibitem[\protect\citeauthoryear{Goel \bgroup \em et al.\egroup
  }{2022}]{goel2022agent}
Ashok~K Goel, Harshvardhan Sikka, and Eric Gregori.
\newblock Agent smith: Machine teaching for building question answering agents.
\newblock In {\em AAAI Spring Symposium: MAKE}, 2022.

\bibitem[\protect\citeauthoryear{Graesser \bgroup \em et al.\egroup
  }{1996}]{graesser1996question}
Arthur~C Graesser, William Baggett, and Kent Williams.
\newblock Question-driven explanatory reasoning.
\newblock {\em Applied Cognitive Psychology}, 10(7):17--31, 1996.

\bibitem[\protect\citeauthoryear{Gunning and Aha}{2019}]{gunning2019darpa}
David Gunning and David Aha.
\newblock Darpa’s explainable artificial intelligence (xai) program.
\newblock {\em AI magazine}, 40(2):44--58, 2019.

\bibitem[\protect\citeauthoryear{Langley \bgroup \em et al.\egroup
  }{2017}]{langley2017explainable}
Pat Langley, Ben Meadows, Mohan Sridharan, and Dongkyu Choi.
\newblock Explainable agency for intelligent autonomous systems.
\newblock In {\em Twenty-Ninth IAAI Conference}, 2017.

\bibitem[\protect\citeauthoryear{Leake and
  Mcsherry}{2005}]{leake2005introduction}
David Leake and David Mcsherry.
\newblock Introduction to the special issue on explanation in case-based
  reasoning.
\newblock {\em The Artificial Intelligence Review}, 24(2):103, 2005.

\bibitem[\protect\citeauthoryear{Lindsay \bgroup \em et al.\egroup
  }{1993}]{lindsay1993dendral}
Robert~K Lindsay, Bruce~G Buchanan, Edward~A Feigenbaum, and Joshua Lederberg.
\newblock Dendral: a case study of the first expert system for scientific
  hypothesis formation.
\newblock {\em Artificial intelligence}, 61(2):209--261, 1993.

\bibitem[\protect\citeauthoryear{Mueller \bgroup \em et al.\egroup
  }{2019}]{mueller2019explanation}
Shane~T Mueller, Robert~R Hoffman, William Clancey, Abigail Emrey, and Gary
  Klein.
\newblock Explanation in human-ai systems: A literature meta-review, synopsis
  of key ideas and publications, and bibliography for explainable ai.
\newblock {\em arXiv preprint arXiv:1902.01876}, 2019.

\bibitem[\protect\citeauthoryear{Muller and
  Druin}{2012}]{muller2012participatory}
Michael~J Muller and Allison Druin.
\newblock Participatory design: the third space in human--computer interaction.
\newblock In {\em The Human--Computer Interaction Handbook}, pages 1125--1153.
  CRC Press, 2012.

\bibitem[\protect\citeauthoryear{Murdock and Goel}{2008}]{murdock2008meta}
J~William Murdock and Ashok~K Goel.
\newblock Meta-case-based reasoning: self-improvement through
  self-understanding.
\newblock {\em Journal of Experimental \& Theoretical Artificial Intelligence},
  20(1):1--36, 2008.

\bibitem[\protect\citeauthoryear{Robson \bgroup \em et al.\egroup
  }{2022a}]{robson2022making}
Robby Robson, Elaine Kelsey, Ashok Goel, Lauren Egerton, Sazzad~M Nasir, Matt
  Lisle, Alan LaFleur, and Elliot Robson.
\newblock Making ai work for skill-based training.
\newblock In {\em International Training Technology Exhibition and Conference},
  2022.

\bibitem[\protect\citeauthoryear{Robson \bgroup \em et al.\egroup
  }{2022b}]{robson2022intelligent}
Robby Robson, Elaine Kelsey, Ashok Goel, Sazzad~M Nasir, Elliot Robson, Myk
  Garn, Matt Lisle, Jeanne Kitchens, Spencer Rugaber, and Fritz Ray.
\newblock Intelligent links: Ai-supported connections between employers and
  colleges.
\newblock {\em AI Magazine}, 43(1):75--82, 2022.

\bibitem[\protect\citeauthoryear{Schank \bgroup \em et al.\egroup
  }{2014}]{schank2014inside}
Roger~C Schank, Alex Kass, and Christopher~K Riesbeck.
\newblock {\em Inside case-based explanation}.
\newblock Psychology Press, 2014.

\bibitem[\protect\citeauthoryear{Spinuzzi}{2005}]{spinuzzi2005methodology}
Clay Spinuzzi.
\newblock The methodology of participatory design.
\newblock {\em Technical communication}, 52(2):163--174, 2005.

\bibitem[\protect\citeauthoryear{USCDC}{2022}]{cdc}
USCDC.
\newblock The importance of ksas.
\newblock \url{https://www.cdc.gov/hrmo/ksahowto.htm}, 2022.
\newblock Accessed: 2022.

\bibitem[\protect\citeauthoryear{USVA}{2022}]{usva}
USVA.
\newblock What are ksas?
\newblock \url{http://www.va.gov/jobs/hiring/apply/ksa.asp}, 2022.
\newblock Accessed: 2022.

\bibitem[\protect\citeauthoryear{Woolf}{2010}]{woolf2010building}
Beverly~Park Woolf.
\newblock {\em Building intelligent interactive tutors: Student-centered
  strategies for revolutionizing e-learning}.
\newblock Morgan Kaufmann, 2010.

\bibitem[\protect\citeauthoryear{Xu \bgroup \em et al.\egroup
  }{2016}]{xu2016question}
Kun Xu, Siva Reddy, Yansong Feng, Songfang Huang, and Dongyan Zhao.
\newblock Question answering on freebase via relation extraction and textual
  evidence.
\newblock {\em arXiv preprint arXiv:1603.00957}, 2016.

\end{thebibliography}

\end{document}